\documentclass[prb,showpacs,twocolumn]{revtex4}
\usepackage{graphics,graphicx}
\begin{document}
\title{Anisotropic itinerant magnetism and spin fluctuations in BaFe$_2$As$_2$: A neutron scattering study}
\author{K.~Matan$^{1,2}$}
 \email{kmatan@issp.u-tokyo.ac.jp}
\author{R.~Morinaga$^{1,2}$}
\author{K.~Iida$^{1}$}
\author{T.~J.~Sato$^{1,2}$}
 \affiliation{$^1$Neutron Science Laboratory, Institute for Solid State Physics, University of Tokyo, 106-1 Shirakata, Tokai, Ibaraki 319-1106, Japan}
 \affiliation{$^2$JST, TRIP, 5, Sanbancho, Chiyoda, Tokyo 102-0075, Japan}
\date{\today}

\begin{abstract}
Neutron scattering measurements were performed to investigate magnetic excitations in a single-crystal sample of the ternary iron arsenide BaFe$_2$As$_2$, a parent compound of a recently discovered family of Fe-based superconductors. In the ordered state, we observe low energy spin-wave excitations with a gap energy $\Delta=9.8(4)$ meV.  The in-plane spin-wave velocity $v_{ab}$ and out-of-plane spin-wave velocity $v_{c}$ measured at 12 meV are 280(150) and 57(7) meV~\AA, respectively. At high energy, we observe anisotropic scattering centered at the antiferromagnetic wave vectors. This scattering indicates two-dimensional spin dynamics, which possibly exist inside the Stoner continuum. At $T_N=136(1)$ K, the gap closes, and quasi-elastic scattering is observed above $T_N$, indicative of short-range spin fluctuations. In the paramagnetic state, the scattering intensity along the $L$ direction becomes ``rodlike,'' characteristic of uncorrelated out-of-plane spins, attesting to the two-dimensionality of the system.
\end{abstract}

\maketitle

\section{Introduction}

A pairing mechanism mediated by magnetic fluctuations has been regarded as a leading candidate to resolve the problem of high-$T_c$ superconductivity. \cite{lee:17} The recent discovery of superconductivity in a family of iron arsenides, \cite{kamihara1} whose non-superconducting parent compounds exhibit long-range antiferromagnetic order similar to the cuprates, \cite{cruz} provides yet another promising group of materials for studying the interplay between magnetism and superconductivity. One property shared by most high-$T_c$ superconductors is two-dimensional magnetism, and both cuprates and iron arsenides are comprised of magnetic layers, sandwiched between non-magnetic ions.  However, while parent compounds of the cuprates are Mott insulators with large in-plane exchange interactions ($J=135$~meV for La$_2$CuO$_4$), \cite{kastner} magnetism in the metallic iron arsenides most likely originates from itinerant electrons, and the antiferromagnetic order is a result of a spin-density-wave (SDW) instability due to Fermi-surface nesting. \cite{mazin:057003} In order to study anisotropic magnetic interactions and itinerant magnetism in the iron arsenides, measurements of magnetic excitations on a single-crystal sample are extremely crucial, and can provide new insights into the understanding of high-$T_c$ superconductivity.

Shortly after the discovery of superconductivity in fluorine-doped LaFeAsO, \cite{kamihara1} Rotter \textit{et al.} \cite{rotter:020503} proposed the ternary iron arsenide BaFe$_2$As$_2$ as a ``very promising candidate'' for superconductivity based on its almost identical structure and electronic properties. Similar to LaFeAsO, BaFe$_2$As$_2$ is comprised of FeAs layers, and shows a transition to an antiferromagnetic ordered state and a structural transition from the tetragonal space group $I4/mmm$ at high temperature to the orthorhombic space group $Fmmm$ at low temperature; though, those transitions appear at different temperatures in the former but at the same temperature in the latter. \cite{cruz,huang} Subsequently, several groups reported superconductivity in potassium, cobalt, cesium, and nickel doped $A$Fe$_2$As$_2$, where $A$=Ba, Sr, Ca, and Eu, with the highest $T_c$ of 38 K in Ba$_{0.6}$K$_{0.4}$Fe$_2$As$_2$. \cite{rotter:107006,sefat:117004, sasmal:107007,jeevan:092406,li:107004}  

Powder neutron diffraction on BaFe$_2$As$_2$ reveals the long-range magnetic order below the N\'eel temperature $T_N=136(1)$ K, where Fe spins with an ordered moment of 0.87(3)~$\mu_B$ per Fe form a colinear spin structure along the crystallographic $a$ axis, and antiferromagnetic (ferromagnetic) arrangement along the $a$ and $c$ axes ($b$ axis). \cite{huang} Previous inelastic neutron scattering studies on a powder sample of BaFe$_2$As$_2$ show magnetic excitations possibly up to 170 meV, indicative of strong spin couplings. \cite{ewing} Neutron scattering studies on single-crystal samples of the related compounds SrFe$_2$As$_2$ ($T_N=220$~K) and CaFe$_2$As$_2$ ($T_N=172$~K) reveal low energy spin-wave excitations with spin gaps of $\leq6.5$ and 6.9(2) meV, respectively. \cite{zhao,mcqueeney} Above the ordering temperature, the spin gaps close, and the spin-wave scattering is replaced by broad quasi-elastic scattering. \cite{mcqueeney}  

\begin{figure}
\centering \vspace{0in}
\includegraphics[width=8.6cm]{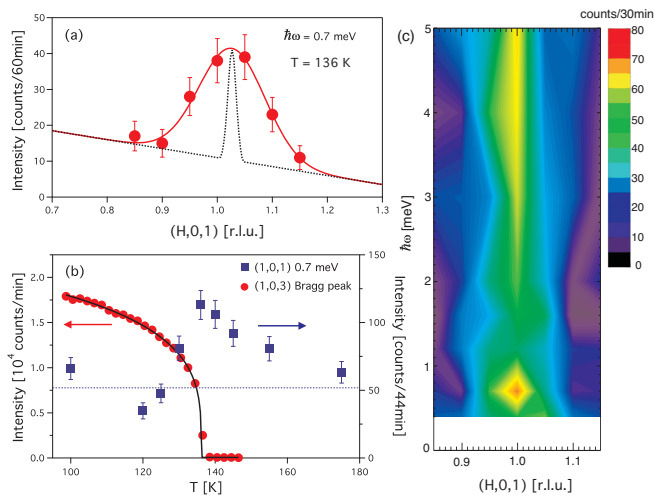}
\caption{(Color online) (a) Constant-energy scans at $\hbar\omega=0.7$ meV and $T=136$ K measured in the flat mode on C1-1. Solid line shows a fit to a Gaussian convoluted with the instrumental resolution function and the dotted line shows the instrumental resolution.  (b) Temperature dependence of the magnetic Bragg intensity measured at (1,0,3) and of magnetic fluctuations measured at $\textbf{Q}=(1,0,1)$ and $\hbar\omega=0.7$ meV using the seven-blade horizontally focused analyzer.  The line corresponds to a power-law form $M^{\dag 2}\propto (T_N-T)^{2\beta}$ for sublattice magnetization $M^\dag$ (which is proportional to the square root of the Bragg intensity) with the exponent $\beta=0.125$ and $T_N=136(1)$ K.  (c) The contour plot as a function energy and $\textbf{Q}$ shows scattering intensity around (1,0,1) at $T=136$ K using the three-blade horizontally focused analyzer.}\label{fig1}
\end{figure}

Here, we present the first inelastic neutron scattering study on a single-crystal sample of BaFe$_2$As$_2$. We observe quasi-elastic scattering, which peaks in the vicinity of $T_N$, indicative of short-range spin fluctuations. Above $T_N$, the out-of-plane spins become uncorrelated and scattering intensity becomes ``rodlike'' along $L$, revealing the two-dimensionality of the system.  Below $T_N$, we observe low energy spin-wave excitations with a spin gap. The difference between in-plane and out-of-plane spin-wave velocities suggests strongly anisotropic magnetism.  Most interestingly, at high energy, we observe anisotropic scattering centered at the antiferromagnetic wave vectors.  This scattering possibly indicates magnetic excitations inside the Stoner continuum.

\section{Experimental details}

A single-crystal sample, which is composed of two coaligned crystals of total mass 0.4~g, was grown using the Bridgman method with the FeAs flux. The detailed synthesis and sample characterizations are reported elsewhere. \cite{morinaga} Neutron-diffraction measurements on this sample yield the ordered moment of $0.91(0.21)~\mu_B$ per Fe, which is consistent with the previous powder-diffraction work. \cite{huang} Inelastic neutron scattering measurements were performed on the triple-axis spectrometers HER (C1-1) (cold neutrons) and GP-TAS (4G) (thermal neutrons), which are operated by the Institute for Solid State Physics, University of Tokyo.  The sample was aligned in the $h0l$ zone with the reciprocal-lattice parameters $a^\ast=1.124(3)$ \AA$^{-1}$ and $c^\ast=0.485(2)$ \AA$^{-1}$. The final energy was fixed at 5 meV at C1-1 and 14.7 meV at 4G.   Vertically focused pyrolytic graphite (PG) crystals were used to monochromate the incident neutron beam using the 002 reflection. Vertically focused (horizontally flat mode) and doubly focused PG crystals were used to analyze the scattered-neutron beam at C1-1, while vertically focused PG crystals were utilized to analyze the scattered-neutron beam at 4G.  At C1-1, horizontal collimations of guide$-$open$-$sample$-80'-80'$ were employed for the flat mode, and guide$-$open$-$sample$-$radial$-$open for the horizontally focusing mode. At 4G, horizontal collimations of $40'-80'-$sample$-80'-80'$ were employed.  PG filters were placed in the scattered beam at 4G, and cooled Be$/$oriented-PG crystals and room-temperature Be filters were placed in the incident and scattered beams at C1-1, respectively, to reduce higher-order contamination. The sample was cooled by a closed cycle $^4$He cryostat.

\section{Spin fluctuations near $T_N$}

A constant-energy scan around (1,0,1) at $\hbar\omega=0.7$ meV and $T=136$ K performed on C1-1 using the horizontally flat analyzer [Fig.~\ref{fig1}(a)] shows spin fluctuations in the vicinity of $T_N$.  The temperature dependence of the low energy scattering at $\textbf{Q}=(1,0,1)$ and $\hbar\omega=0.7$ meV using the seven-blade horizontally focused analyzer [Fig.~\ref{fig1}(b)] shows a peak at $T_N$, which coincides with the onset of the magnetic Bragg intensity measured at (1,0,3).  As a function of temperature, this scattering intensity rapidly decreases on the low-temperature side, resulting from opening of the spin gap (the detailed discussion of the spin gap will be presented below), but has much weaker temperature dependence on the high-temperature side, indicative of critical scattering above $T_N$.  Several constant-energy scans performed at 136 K using the three-blade horizontally focused analyzer constitute a contour plot [Fig.~\ref{fig1}(c)], which shows a non-dispersive scattering column centered at the antiferromagnetic Brillouin-zone center (1,0,1) that appears to extend up to high energy.  The scattering column, which exists at least up to 2 meV (the highest energy of the C1-1 measurements), was also observed around (1,0,3) at 136 K (not shown). The constant-energy scans at $\hbar\omega=0.7$ meV measured at C1-1 [Fig.~\ref{fig1}(a)] and at $\hbar\omega=12$ meV measured at 4G (not shown) yield correlation lengths $\xi=15(1)$ \AA~and $\xi=18(2)$ \AA, respectively, corresponding to roughly 6 times the nearest-neighbor distance. 

Our neutron scattering measurements show that the scattering near $T_N$ are dominated by short-range spin fluctuations. The presence of these short-range fluctuations might appear to contradict a report of the first-order magnetic transition observed in the $^{75}$As nuclear-magnetic-resonance study. \cite{kitagawa}  However, large magnetic interactions can give rise to the short-range spin fluctuations in the paramagnetic state, despite the first-order magnetic transition. In addition, two-dimensionality in this layered system can enhance spin fluctuations. The temperature dependence of the order parameter measured with neutron scattering shows no sign of hysteresis. Furthermore, the spin gap decreases continuously as a function of temperature (not shown), showing no sign of an abrupt change at $T_N$ that is expected for a first-order transition. On the other hand, we do not observe the divergence of correlation length at $T_N$, which would suggest a second-order transition. Therefore, the nature of the magnetic transition in BaFe$_2$As$_2$ cannot be resolved by our neutron scattering measurements. However, neutron scattering measurements of the order parameter on a related compound $\alpha$-FeTe show thermal hysteresis at the transition, supporting the first-order magnetic transition. \cite{bao}

\begin{figure}
\centering \vspace{0in}
\includegraphics[width=8.6cm]{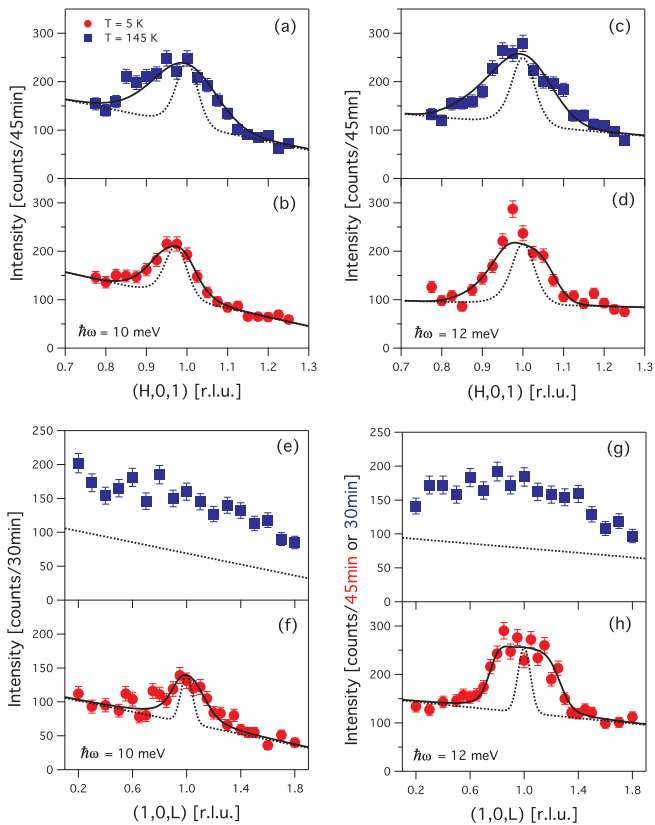}
\caption{(Color online) Constant-energy scans around (1,0,1) at $\hbar\omega=10$ and 12~meV, and at $T=5$ and 145~K. (a)$-$(d) show the scans along $H$ and (e)$-$(h) along $L$.  Solid lines in (b), (d), (f), and (h) show fits to the empirical dispersion relation described in the text convoluted with the instrumental resolution function.  Solid lines represent the best fits with $\Delta=9.8(4)$ meV, $v_{ab}=280(150)$ meV~\AA,~and $v_{c}=57(7)$ meV~\AA. The $L$ scans at $T=145$ K, (a) and (b), are fit with Gaussians convoluted with the resolution function. Dotted lines in (a)$-$(d) indicate the instrumental resolution, while the dotted lines in (e) and (g) show the background level estimated from the low-temperature data.  }\label{fig2}
\end{figure} 

\section{Low energy spin-wave excitations}

In the ordered state, scattering from low energy spin-wave excitations was observed.  Figures~\ref{fig2}(b), 2(d), 2(f), and 2(g) show a series of constant-energy scans around (1,0,1) measured at 4G at $T=5$ K.  The observed peaks were fit with narrow (resolution-limited) Lorentzians $I_{sw}({\bf q},\hbar\omega)=\frac{A(\hbar\omega)}{\pi}\frac{\hbar\Gamma}{(\hbar\Gamma)^2+(\hbar\omega-\hbar\omega_{\bf q})^2}$ convoluted with the instrumental resolution function taking into account the following empirical spin-wave dispersion relation:
\begin{equation}
\hbar\omega_{\bf q}=\sqrt{\Delta^2+v_{ab}^2(q_x^2+q_y^2)+\left(v_cq_z\right)^2},\label{disp}
\end{equation}
\noindent where $\Delta$ is a gap energy in meV, and $v_{ab}$ and $v_c$ are in-plane and out-of-plane spin-wave velocities in a unit of meV~\AA, respectively. ${\bf q}=(q_x, q_y, q_z)$ is a wave vector in the first Brillouin zone away from the zone center {\bf Q$_{\bf A}$} in a unit of \AA$^{-1}$.  It should be noted that this spin-wave model is used as a tool to analyze the data, and does not imply a localized spin model. $\Delta$ was obtained by fitting the 10 meV constant-energy scans, since these scans were performed in the proximity of the spin gap and, therefore, are the most sensitive to $\Delta$.  $v_{ab}$ and $v_{c}$ were obtained by fitting the 12 meV constant-energy scans. Using the fit parameters from the previous round of fitting as inputs, we repeated the fitting several times to assure that all fit parameters are self-consistent, since the three parameters are not independent due to a finite instrumental resolution. The obtained fit parameters are $\Delta=9.8(4)$ meV, $v_{ab}=280(150)$ meV~\AA,~and $v_{c}=57(7)$ meV~\AA, where the error corresponds to three times the statistical error. The gap energy is consistent with a constant-$\textbf{Q}$ scan measured at (1,0,3) [Fig.~\ref{fig3}(a)]. The overall prefactor $A(\hbar\omega)$ follows the $1/\hbar\omega$ scaling within the error bars, consistent with a report in Ref.~\onlinecite{mcqueeney}. The black solid lines in Fig.~\ref{fig2} show the best fits to the empirical dispersion relation [Eq.~\ref{disp}] using the aforementioned parameters. The line shape and apparent shift from the Brillouin-zone center are governed by the four-dimensional resolution function. The dotted lines indicate the instrumental resolution, assuming a $\delta$-function excitation. The spin-wave velocities are expected to increase at high energy away form the spin gap; Fig.~\ref{fig3}d shows $v_{c}=80(10)$ meV~\AA~at $\hbar\omega=24$ meV.  Unfortunately, as will be discussed below, a scan along $H$ at high energy does not allow accurate determination of $v_{ab}$.

As a comparison, the spin-wave velocities for CaFe$_2$As$_2$ are $v_{ab}=420(70)$ meV~\AA~and $v_{c}=270(100)$ meV~\AA, \cite{mcqueeney} and for SrFe$_2$As$_2$,~$v_{ab}=560$ meV~\AA~and $v_{c}=280$ meV~\AA~(note a factor-of-2 difference from Ref. \onlinecite{zhao}).  Compared to CaFe$_2$As$_2$ and SrFe$_2$As$_2$, the much smaller $v_{c}$ suggests the significantly weaker interlayer coupling, and hence lower $T_N$ in BaFe$_2$As$_2$. Furthermore, the smaller ratio of $v_{c}$ to $v_{ab}$ indicates the more anisotropic magnetic interactions in this system.  As a further comparison, the spin gap in BaFe$_2$As$_2$ is larger than those in CaFe$_2$As$_2$ and SrFe$_2$As$_2$. The origin of the spin gaps in this family of FeAs compounds is still under much debate. In other itinerant antiferromagnetic systems, for example, in a family of $\gamma$-Mn alloys, where spin gaps of 7-10 meV were observed, \cite{tajima,mikke1,mikke2} Fishman and Liu \cite{fishman} have successfully derived the temperature dependence of the spin gaps in agreement with the experiments using a two-band model with magnetoelastic interactions. In the iron arsenide systems, the closeness of the structural and magnetic phase transitions, which has also been observed in the $\gamma$-Mn alloys, provides tantalizing evidence for strong magnetoelastic interactions that could explain the presence of the spin gaps.  In addition, dipole interactions and spin-orbit couplings could also contribute to the presence of the spin gap in these systems.

At $T=145$ K above $T_N$, the spin-wave scattering is replaced by broad quasi-elastic scattering that extends up to high energy. The $H$ scans in Figs.~\ref{fig2}(a) and 2(c) yield correlation lengths $\xi=13(2)$ \AA~at $\hbar\omega=10$ meV, and $\xi=15(2)$ \AA~at $\hbar\omega=12$ meV, which are slightly smaller at this temperature compared to 136 K as the correlation lengths are expected to decrease as temperature increases.  On the other hand, the $L$ scans in Figs.~\ref{fig2}(e) and 2(g) show rodlike scattering, indicative of uncorrelated interlayer spins.  The two-dimensionality of this system is clearly observed in the paramagnetic state with large in-plane correlation lengths and uncorrelated out-of-plane spins.  Similar to the cuprates, the relatively large correlation lengths in the paramagnetic state are a consequence of the large magnetic interactions and the two-dimensionality of the magnetism. \cite{kastner} Despite this similarity, however, the origin of magnetism in the iron arsenides is believed to be quite different from the cuprates.

\section{Magnetic excitations at high energy}

\begin{figure}
\centering \vspace{0in}
\includegraphics[width=8.6cm]{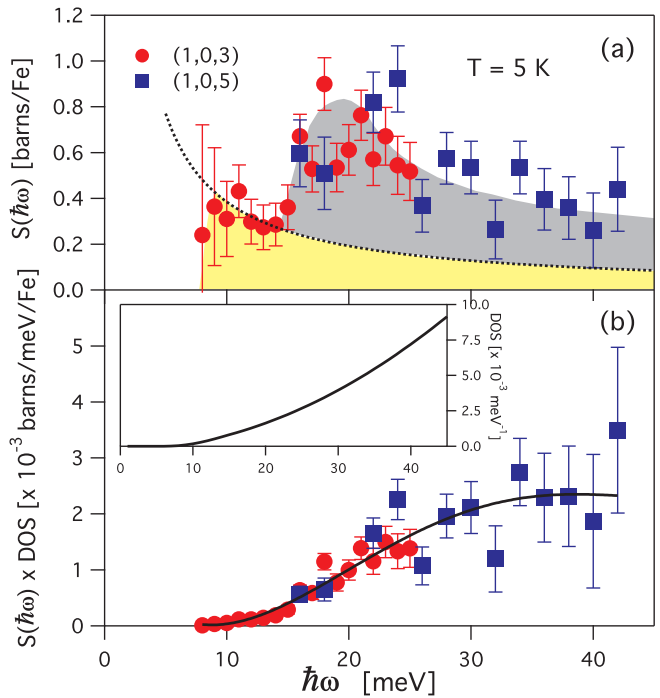}
\caption{(Color online) (a) Scattering intensity in barn/meV at $T = 5$ K measured at (1,0,3) (red circles) and (1,0,5) (blue squares) after corrected for background, absorption, the instrumental resolution, and the magnetic form factor. Dotted line shows the $1/\hbar\omega$ dependence. Shaded areas are guides to the eyes, representing the spin-wave scattering (yellow) and the ``extra'' scattering (gray).  (b) shows a product of $S(\hbar\omega_{\bf q})$ and the density of states. Solid line is a guide to the eyes. Inset shows the density of states.}\label{fig3}
\end{figure}

In order to further investigate the nature of magnetism in BaFe$_2$As$_2$, we measure inelastic scattering in an absolute unit of barn/meV at $T = 5$ K to estimate a fluctuating moment.  A normalization function to inelastic scattering was obtained from acoustic phonons measured on copper.  The normalized intensity can be described by the following equation:
\begin{equation}
I({\bf q},\hbar\omega)=\left(\frac{\gamma r_0}{2}\right)^2\left[f({\bf Q})\right]^2\widetilde{S}(\hbar\omega)\delta(\hbar\omega-\hbar\omega_{\bf q}),
\end{equation}
where $\left(\frac{\gamma r_0}{2}\right)^2=72.65\times10^{-3}$ barn$/\mu_B^2$, $f({\bf Q})$ is a magnetic form factor for Fe$^{2+}$, and $\hbar\omega_{\bf q}$ is the spin-wave energy described by the dispersion relation in Eq.~\ref{disp}. Figure~\ref{fig3}(a) shows the scattering intensity $S(\hbar\omega_{\bf q})=\left(\frac{\gamma r_0}{2}\right)^2\widetilde{S}(\hbar\omega_{\bf q})$ in barn/Fe measured at ${\bf Q} = (1,0,3)$ and (1,0,5) after corrected for background, absorption, the instrumental resolution, and the magnetic form factor.  The same scattering was also observed at the antiferromagnetic wave vectors (3,0,1) and (1,0,7), whose different intensities can be accounted for by the magnetic form factor and polarization effect.  To obtain the fluctuating moment $M_f$,  we calculate 
\begin{eqnarray}
M_f^2=\left(\frac{2}{\gamma r_0}\right)^2\int S(\hbar\omega)\delta(\hbar\omega-\hbar\omega_{\bf q})~d{\bf q}~d\omega.\nonumber
\end{eqnarray}
Since the ordered moment is along the crystallographic $a$ axis, nearly equivalent intensities at (3,0,1) and (1,0,3) suggest that $M_f$ are dominated by transverse fluctuations. Figure~\ref{fig3}(b) shows a product of $S(\hbar\omega_{\bf q})$ and the density of states, assuming the dispersion relation in Eq.~\ref{disp}.  The integral was calculated up to 42 meV.  We obtain $M_f=0.8(2)~\mu_B$, which already accounts for most of the ordered moment. However, the magnetic excitations were reported to extend up to much higher energy. \cite{ewing} This discrepancy suggests magnetic scattering from other origin.

\begin{figure}
\centering \vspace{0in}
\includegraphics[width=8.6cm]{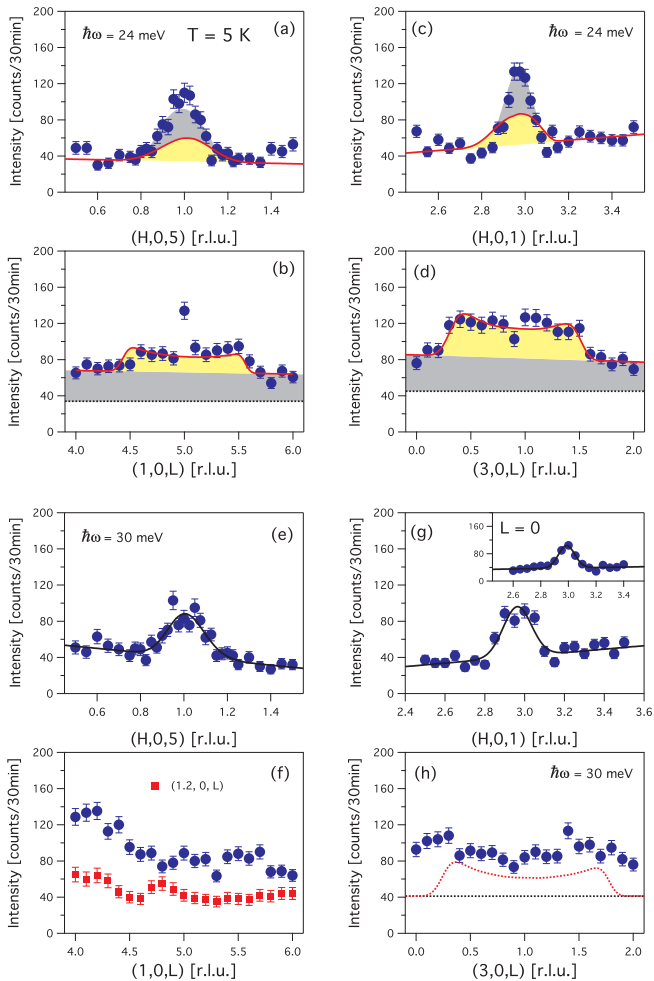}
\caption{(Color online) Constant-energy scans at 24 meV around [(a)$-$(b)] (1,0,5) and [(c)$-$(d)] (3,0,1). Solid lines show the spin-wave scattering with $\Delta=9.8(4)$ meV, $v_{ab}=280$ meV~\AA, and $v_{c}=80$ meV~\AA.  Dotted line in (b) and (d) shows background estimated from the $H$ scans.  Shaded areas, which serve as guides to the eyes, show the spin-wave scattering (yellow), and the extra scattering (gray).  (e) and (f), and (g) and (h) show constant-energy scans at 30 meV around (1,0,5) and (3,0,1), respectively.  The inset in (g) shows a $H$ scan at $L=0$.  Black solid lines in (e) and (g) are guides to the eyes. Red squares in (f) indicate background measured around (1.2,0,5).  The rise on the low-$q$ side is due to the main beam.  The red dotted line in (h) shows the spin-wave scattering with the above parameters. All data were measured with a counting time of 30 min or higher, but are normalized to counts/30min.}\label{fig4}
\end{figure}

In fact, constant-energy scans at $\hbar\omega=24$ meV show an ``extra'' scattering intensity, which is centered at the antiferromagnetic wave vectors and coexisting with the spin-wave scattering. The gray shaded areas in Figs.~\ref{fig4}(a)$-$4(d) show this scattering around (1,0,5) and (3,0,1), which is sharp along $H$, but broad along $L$. The yellow shaded areas indicate the spin-wave scattering with $\Delta=9.8(4)$ meV, $v_{ab}=280$ meV~\AA, and $v_{c}=80$ meV~\AA; note a larger $v_{c}$ at this energy.  Unfortunately, the presence of the extra scattering in the $H$ scans hinders accurate determination of $v_{ab}$, which presumably becomes larger than 280 meV~\AA~at this energy. Consequently, the shaded areas in Figs.~\ref{fig4}(a) and 4(c) are rough estimates and only serve as guides to the eyes. Figures~\ref{fig4}(b) and 4(d) describe the scattering intensities along $L$, which clearly show the coexistence of the spin-wave scattering and the extra scattering. The difference of the spin-wave scattering prefactor $A(\hbar\omega)$ at (1,0,5) and (3,0,1) can be explained by the transverse spin waves given the ordered moment along $a$. A single high point in Fig.~\ref{fig4}(b) is spurious due to neutrons scattered off the aluminum frame of the analyzer.  It was not observed at any other antiferromagnetic wave vectors or energies and hence has a negligible effect on $M_f$. At $\hbar\omega=30$ meV [Figs.~\ref{fig4}(e) and 4(f)), the spin-wave scattering disappears or becomes negligibly small and the magnetic excitations are dominated by the extra scattering. Figures~\ref{fig4}(e) and 4(g) show sharp peaks along $H$ around (1,0,5) and (3,0,1), which are also observed at a different $L$ as shown in the inset for $L=0$. On the other hand, Figs.~\ref{fig4}(f) and 4(h) show very broad scattering along $L$. As shown in Fig.~\ref{fig4}(h), this broadening along $L$ cannot be accounted for by the spin-wave scattering. This extra scattering exhibits an anisotropic character with the broad $L$ dependence and sharp $H$ dependence, suggesting two-dimensional spin dynamics.

This anisotropic scattering at high energy is possibly indicative of electron-hole excitations inside the Stoner continuum. In Fig.~\ref{fig3}(a), we observe the rise of $S(\hbar\omega)$ starting at 15 meV and reaching a maximum around 20 meV (shown by the gray shaded area), which deviates from the $1/\hbar\omega$ dependence shown by the dotted line. According to a weak-coupling theory, \cite{fedders} an energy gap, above which the collective spin-wave excitations are expected to merge into the Stoner electron-hole continuum, is given by $\Delta_s=1.76\alpha k_BT_N$, where $\alpha=(v_1+v_2)/(4v_1v_2)^{1/2}$; $v_1$ and $v_2$ are Fermi velocities of electrons and holes, respectively.  For BaFe$_2$As$_2$, $v_1\sim v_2\sim0.5$ eV~\AA, \cite{zhao} and therefore $\Delta_s\sim20$~meV.  As a comparison, assuming the same Fermi velocities, $\Delta_s$'s are 26 and 33 meV for CaFe$_2$As$_2$ and SrFe$_2$As$_2$, respectively. The larger $\Delta_s$'s might explain why this anisotropic scattering has not yet been observed in CaFe$_2$As$_2$ and SrFe$_2$As$_2$. \cite{zhao,mcqueeney} Similar broad inelastic-scattering lobes, which represent magnetic excitations inside the Stoner continuum, have been observed in V$_{2-\textnormal{y}}$O$_3$ ($\Delta_s=1.4$~meV) around incommensurate Bragg peaks below $T_N$. \cite{baovo}  In BaFe$_2$As$_2$, we possibly observe the coexistence of the collective spin-wave excitations and electron-hole excitations near the edge of the Stoner continuum at 24 meV and only the single-particle spin dynamics above 30 meV.  However, the presence of the spin-wave scattering inside the Stoner continuum and the anisotropic character appear to differ from V$_{2-\textnormal{y}}$O$_3$.  Further experimental work on BaFe$_2$As$_2$, as well as on other iron arsenide compounds, is needed to further study these two-dimensional spin dynamics at higher energy.

\section{Conclusion}

The magnetic excitations in BaFe$_2$As$_2$  have been measured using inelastic neutron scattering.  We observe low energy spin-wave excitations with a spin gap in the ordered state.  The smaller ratio of the out-of-plane to in-plane spin-wave velocities compared to CaFe$_2$As$_2$ and SrFe$_2$As$_2$ suggests that the magnetism in BaFe$_2$As$_2$ is more two-dimensional. At high energy, we observe the anisotropic scattering centered at the antiferromagnetic wave vectors. This scattering possibly indicates two-dimensional spin dynamics inside the Stoner electron-hole continuum. In the vicinity of $T_N$, the spin gap closes, and the spin fluctuations are observed up to 12 meV (the highest energy of our measurements).  Above $T_N$, these spin fluctuations show the anisotropic character, attesting to the two-dimensionality of this system. It is believed that two-dimensionality plays a significant role in high-$T_c$ superconductivity. The fact that potassium-doped BaFe$_2$As$_2$ has the highest $T_c$ so far among the nonoxide iron arsenide compounds hints at the connection between superconductivity and two-dimensional magnetism. It is interesting to see if two-dimensional spin fluctuations exist in the superconducting state.

\begin{acknowledgments}
We thank M.~Ogata, H.~Yoshizawa, K.~Ohgushi, S.-H.~Lee, M.~Nishi, H.~Suzuki, and T.~G.~Perring for useful discussions.  We also thank R.~Akiyama and S.~Ibuka for their technical helps.
\end{acknowledgments}

\bibliography{BaFe2As2-neutron}

\end{document}